\documentclass[11pt,lettersize]{article}
\usepackage{sober}
\usepackage{amsthm,amsmath, amssymb}
\usepackage{color,graphicx,subfigure,geometry,boxedminipage}
\usepackage{threeparttable}
\usepackage{bbding}
\usepackage{hyperref,verbatim}
\usepackage[noend]{algorithmic}
\usepackage[section,boxed]{algorithm}
\usepackage{array}
\usepackage{afterpage}

 \title{Self-healing systems and virtual structures}
\author{ Amitabh Trehan
\thanks{Information Systems Group, Faculty of Industrial Engineering and Management, Technion, Haifa, Isreal. Supported by a Technion Fellowship. email: {\tt amitabh.trehaan@gmail.com}}
\date{}
}

\newtheorem{observation}{Observation}

\DeclareGraphicsExtensions{.pdf}

\newcommand{\RT}{\mathrm{RT}}

\newcommand{\RStr}{\mathrm{Reconstruction Structure}}

\newcommand{\Degree}{\mathrm{degree}}

\newcommand{\G}{\mathrm{G}}

\renewcommand{\algorithmiccomment}[1]{// #1} 
\begin{document}

\maketitle

\begin{abstract}

Modern networks are large, highly complex and dynamic. Add to that the mobility of the agents comprising many of these networks. It is difficult or even impossible for such systems to be managed centrally in an efficient manner. It is imperative for such systems to attain a degree of self-management.  Self-healing i.e. the capability of a system in a good state to recover to another good state in face of an attack, is desirable for such systems. 
In this paper, we discuss the self-healing model for dynamic reconfigurable systems. In this model, an omniscient adversary inserts or deletes nodes from a network and the algorithm responds by adding a limited number of edges in order to maintain invariants of the network. 
We look at some of the results in this model and argue for their applicability and further extensions of the results and the model. We also look at some of the techniques we have used in our earlier work, in particular, we look at the idea of maintaining virtual graphs mapped over the existing network and assert that this may be a useful technique to use in many problem domains.
\end{abstract}

\section{Introduction}
\label{sec: intro}

Modern networks have evolved to become  both large and highly complex, with some networks spanning nations and even the globe. Networks provide a multitude of services using a wide variety of protocols and  components to the extent that they have now begun to resemble self-governed living entities. 
Most modern networks are dynamic with nodes entering the network or leaving by choice, failure or attack. There are dynamic networks which have always been around in some form, like social networks, which we have only now begun to analyze and in fact, influence. That maintaining robustness in modern networks can be an issue can be ascertained by the regular breakdowns in large and important networks e.g. the crash of the Skype network in 2007~\cite{fisher,malik, moore, ray, stone} attributed to the failure of its ``self-healing'' mechanisms~\cite{garvey}.
Also, due to the scale and nature of design of such networks, it may simply not be practical to build robustness into the individual nodes or into the structure of the initial network itself. Thus, the need for a responsive approach to robustness. Many important networks are also \emph{reconfigurable}  in the sense that they can change their topology e.g. peer-to-peer, wireless, ad-hoc networks and friendship networks on social networking sites etc.\ . We exploit this property of networks to allow us a responsive approach towards robustness. Moreover, our algorithms are scalable since our repair costs are constant or at most logarithmic in the number of nodes, and inherently handle the dynamism of the network. Also, we conjecture that some of the techniques we use, in particular, virtual graphs can be effectively used for a wider range of problems than we discuss.

Informally,  self-healing is the maintenance of certain properties within desirable bounds by the nodes in a network suffering from failures or under attack. As the name implies, self-healing has to be initiated and executed by the nodes themselves. As such, the self-healing algorithms we have devised are fully distributed. We can say  that a self-healing system, when starting from a correct state, can only be temporarily out of a  correct state i.e. it recovers to a correct state, in presence of  attacks. 


Our sense of self-healing is more formally captured by the model discussed in Section~\ref{sec: Intro-self-healingModel}. 
Informally, the model we adopt in this work is as follows. We assume (for simplicity of description) that the network is initially a connected graph over $n$ nodes.  An adversary repeatedly attacks the network. This adversary knows the network topology and our algorithm, and it has the ability to delete arbitrary nodes from the  network or insert a new node in the system which it can connect to any subset of the nodes currently in the system.   However, we assume the adversary is constrained in that in any time step it can only delete or insert a single node. Following that, the self-healing algorithm has a short time to reconfigure and heal the network by adding edges between remaining nodes before the next act of the adversary. Our model  could, for example, capture what can happen when a worm or software error propagates through the population of nodes. We have developed a series of self-healing algorithms: \emph{DASH}~\cite{SaiaTrehanIPDPS08}, \emph{ForgivingTree}~\cite{HayesPODC08}, \emph{ForgivingGraph}~\cite{FG-DCJournal2012,HayesPODC09}, \emph{Xheal}~\cite{PanduranganPODC11},\emph{Xheal+}~\cite{SarmaTrehanEdgeNetSciComm2012} that we succintly compare in Section~\ref{sec: reconstructionstructs}.
Though our algorithms are directly applicable to reconfigurable computer networks, the notion of self-healing is  important across different domains.

\medskip
\noindent {\bf Our Contributions:} 
In this paper, we contend that: a) The self-healing model is a powerful and flexible model to study and design reconfigurable dynamic networks, and b) We introduce virtual graphs (and suggest a framework) contending that they are powerful tools e.g. for designing self-healing solutions.

\medskip
\noindent {\bf Related Work:} 
Self-healing is one of the so called `Self-*' properties which systems such as autonomic systems~\cite{IBMAutonomicVision} may be required to have.  In the distributed systems world, perhaps the most well-known self-* property is \emph{self-stabilization}~\cite{Djikstra74SelfStabilizing, DolevBookSelfStabilization, Dolev09EmpireofColoniesSelf-stabilizing, GerardTelDistributedAlgosBook}. Self-stabilization was introduced by Djikstra in 1974~\cite{Djikstra74SelfStabilizing}. A self-stabilizing system is a system  which, starting from an arbitrary state and being affected by adversarial transient failures, can, in finite time, recover to a correct state.
  Other self-* properties, often broadly defined, include  \emph{self-scaling, self-repairing} (similar to self-healing), \emph{self-adjusting} (similar to self-managing), \emph{self-aware/self-monitoring, self-immune, self-containing}~\cite{Berns09DissectingSelf-*}.

Self-healing is a responsive approach to reliable systems. 
  There have been numerous other papers that
discuss strategies for adding additional capacity or rerouting in
anticipation of failures \cite{doverspike94capacity,
frisanco97capacity, iraschko98capacity, murakami97comparative,
caenegem97capacity, xiong99restore}.  Results that
are responsive in some sense include the following.  M\'{e}dard, Finn, Barry, and Gallager
\cite{medard99redundant} propose constructing redundant trees to make
backup routes possible when an edge or node is deleted.  Anderson,
Balakrishnan, Kaashoek, and Morris \cite{anderson01RON} modify some
existing nodes to be RON (Resilient Overlay Network) nodes to detect
failures and reroute accordingly. Some networks have enough redundancy
built in so that separate parts of the network can function on their
own in case of an attack~\cite{goel04resilient}.  In all these past
results, the network topology is fixed.  In contrast, our approach
adds edges to the network as node failures occur.  Further, our
approach does not dictate routing paths or specifically require
redundant components to be placed in the network initially.   

Dynamic network topology and fault tolerance have always been core concerns of distributed computing~\cite{Attiya-WelchBook,Lynchbook}. There are many models and a large volume of work in this area. The self-healing model is a suitable model for overlay networks in a dynamic setting. Broadly, dynamic models may be classified as node-dynamic or edge-dynamic. Some reconfigurable overlay network based models are node-dynamic in that nodes join and leave continously~\cite{Kuhn2005Self-Repairing,Amitabh-2010-PhdThesis}. A special class of this is the self-healing model where the algorithm can add a limited number of edges in response to a deletion~\cite{Poor-SelfHealQueue2003, Ghosh07Self-healingSystemsSurvey}.
A notable recent edge-dynamic model is the dynamic graph model introduced by Kuhn, Lynch and Oshman in ~\cite{Kuhn-DistComputation-STOC10}. They introduced a  stability property called T-interval connectivity (for $T\ge 1$) which stipulates the existence of a stable connected spanning subgraph for every $T$ rounds.


There has also been research in the physics community on
preventing cascading failures.  In the model used for these results,
each vertex in the network starts with a fixed capacity. When a vertex
is deleted, some of its ``load'' (typically defined as the number of
shortest paths that go through the vertex) is diverted to the
remaining vertices.  The remaining vertices, in turn, can fail if the
extra load exceeds their capacities. Motter, Lai, Holme, and Kim have
shown empirically that even a single node deletion can cause a
constant fraction of the nodes to fail in a power-law network due to
cascading failures\cite{holme-2002-65, motter-2002-66}. Motter and Lai
propose a strategy for addressing this problem by intentional removal
of certain nodes in the network after a failure begins
\cite{motter-2004-93}.  Hayashi and Miyazaki propose another strategy,
called emergent rewirings, that adds edges to the network after a
failure begins to prevent the failure from
cascading\cite{hayashi2005}.  Both of these approaches are
shown to work well empirically on many networks.  However, unfortunately, they
perform very poorly under adversarial attack.


 \subsection{Model of self-healing}
\label{sec: Intro-self-healingModel}


Our general model of self-healing is shown in Figure~\ref{algo: model-general}. This model was introduced in~\cite{Amitabh-2010-PhdThesis}. It is generalized from the model in~\cite{FG-DCJournal2012, HayesPODC09}. Somewhat similar models were also used in~\cite{ PanduranganPODC11, SarmaTrehanEdgeNetSciComm2012, HayesPODC08,SaiaTrehanIPDPS08}.
The specific models used in most of our algorithms are special cases of this model, differing mainly in the way the success metrics of the graph properties are presented. The model used in Xheal~\cite{PanduranganPODC11, SarmaTrehanEdgeNetSciComm2012} also differs in the synchronicity and message assumptions.

Let $G = G_0$ be an arbitrary graph on $n$ nodes, which represent processors in a distributed network.  In each step, the adversary either deletes or adds a node.  After each deletion, the algorithm gets to add some new edges to the graph, as well as deleting old ones.  At each insertion, the processors follow a protocol to update their information.
The algorithm's goal is to maintain the chosen graph properties within the desired bounds. At the same time, the algorithm wants to minimize the resources spent on this task.  Initially, each processor only knows its neighbors in $G_0$, and is unaware of the structure of the rest of $G_0$. After each deletion or insertion, only the neighbors of the deleted or inserted vertex are informed that the deletion or insertion has occured. After this, processors are allowed to communicate by sending a limited number of messages to their direct  neighbors.  We assume that these messages are always sent and received successfully.  The processors may also request new edges be added to the graph.  

\begin{figure}[h!]
\centering
\subfigure[$G'_T$: Nodes in red (dark gray in grayscale) deleted, and nodes in green (patterned) inserted, by the adversary.]{ \label{sfig: GraphOrigInserts}
 \makebox[0.4\textwidth][c]{\includegraphics[scale=0.6]{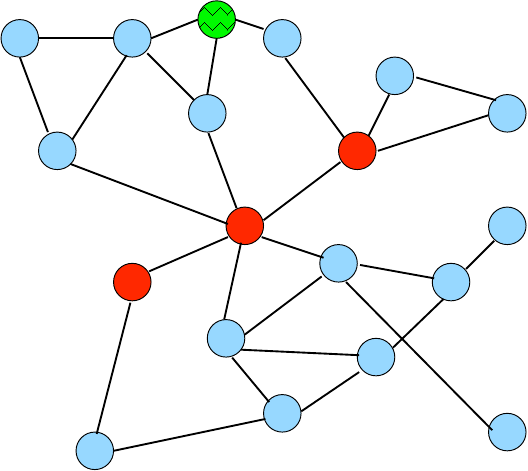}} }
\hspace{0.1 \textwidth}
\subfigure[$G_T$: The actual graph. Edges added by the healing algorithm shown in gold (light shaded in grayscale) color.]
{ \label{sfig: GraphHealed}
\makebox[0.4\textwidth][c]{ \includegraphics[scale=0.6]{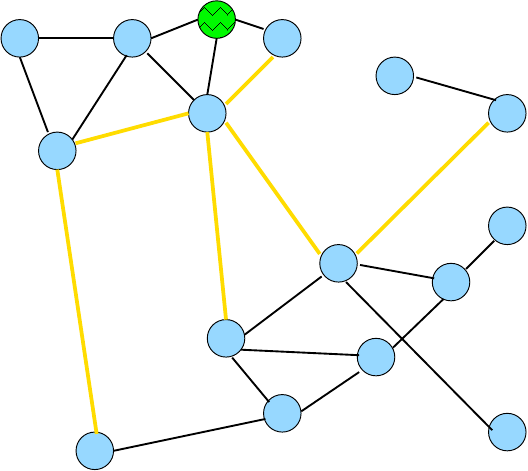} }}
\caption{ \emph{Graphs at time T}. $G'_T$: The graph of initial nodes and insertions over time, $G_T$: The actual healed graph.}
\label{fig: Intro-ComparisonGraphs}
\end{figure}

We also allow a certain amount of pre-processing to be done before the first attack occurs.  This may, for instance,
be used by the processors to gather some topological information about $G_0$, or perhaps to 
coordinate a strategy.  Another success metric is the amount of computation and communication needed during this
preprocessing round.  For our success metrics, we compare the graphs at time $T$: the actual graph $G_T$ to the graph $G'_{T}$ which is the graph with only the original nodes (those at $G_0$) and insertions without regard to deletions and healing. This is the graph which would have been present if the adversary was not doing any deletions and (thus) no self-healing algorithm was active. This is the natural graph for comparing results.  
Figure~\ref{fig: Intro-ComparisonGraphs} shows an example of $G'_T$ and a corresponding $G_T$. The figure also shows, in  $G'_T$,  the nodes and edges inserted  and deleted,  and in $G_T$, the edges inserted by the healing algorithm, as the network evolved over time. 

\begin{figure}[h!]
\caption{The general distributed Node Insert, Delete and Network Repair Model.}
\label{algo: model-general}
\begin{boxedminipage}{\textwidth}
\begin{algorithmic}
\STATE Each node of $G_0$ is a processor.  
\STATE Each processor starts with a list of its neighbors in $G_0$.
\STATE Pre-processing: Processors may exchange messages with their neighbors.
\FOR {$t := 1$ to $T$}
\STATE Adversary deletes a node $v_t$ from $G_{t-1}$ or inserts a node $v_t$ into $G_{t-1}$, forming $H_t$.
\IF{node $v_t$ is inserted} 
\STATE The new neighbors of $v_t$ may update their information and exchange messages with their neighbors.
\ENDIF
\IF{node $v_t$ is deleted} 
\STATE All neighbors of $v_t$ are informed of the deletion.
\STATE {\bf Recovery phase:}
\STATE Nodes of $H_t$ may communicate (asynchronously/synchronously, in parallel) 
with their immediate neighbors.  These messages are never lost or
corrupted, and may contain the names of other vertices.
\STATE During this phase, each node may add edges
joining it to any other nodes as desired. 
Nodes may also drop edges from previous rounds if no longer required.
\ENDIF
\STATE At the end of this phase, we call the graph $G_t$.
\ENDFOR
\vspace{5pt}
\hrule
\STATE
\STATE {\bf Success metrics:} Minimize the following ``complexity'' measures:\\
Consider the graph  $G'$ which is the graph consisting solely of the original nodes and insertions without regard to
deletions and healings. Graph $G'_{t}$ is $G'$ at timestep $t$ (i.e. after the $t^{\mathrm{th}}$ insertion or deletion).
 \begin{enumerate}
\item{\bf Graph properties/invariants.}\\
The graph properties/ invariants we are trying to preserve. e.g.  \emph{Degree increase:} 
  $\max_{v \in G} \Degree(v,G_T) / \Degree(v,G'_T)$
\item{\bf Communication per node.} The maximum number of bits sent by a single node in a single recovery round.
\item{\bf Recovery time.} The maximum total time for a recovery round,
assuming it takes a message no more than $1$ time unit to traverse any edge and we have unlimited local computational power at each node.
\end{enumerate}
\end{algorithmic}
\end{boxedminipage}
\end{figure}


\section{The idea of Reconstruction Structures}
\label{sec: reconstructionstructs}

\begin{table}[h!]
\centering
\begin{threeparttable}[b]
\begin{tabular}{|l|c|c|p{0.08\textwidth}|p{0.12\textwidth}|p{0.11\textwidth}|p{0.08\textwidth}|}
\hline
 & \multicolumn{2}{|c|}{Adversarial Attack} &  \multicolumn{4}{|c|}{Property bounded}\\
 \hline
 & Deletion & Insertion & Connec\-tivity & Degree \mbox{(orig: d)\tnote{\textasteriskcentered}} & Diameter \mbox{(orig: D)\tnote{\textasteriskcentered}} & Stretch \\
 \hline
 DASH & \checkmark & \checkmark& \checkmark &$d + 2\log n$ & \textemdash & \textemdash \\
 \hline
 Forgiving Tree\tnote{!} & \checkmark&$\times $& \checkmark & $d + 3$ & $D\log \Delta$  & \textemdash\\
 \hline
  Forgiving Graph\tnote{!} & \checkmark& \checkmark & \checkmark & $3d$ & $D\log n$ & $\log n$ \\
  \hline
  Xheal/Xheal+ & \checkmark& \checkmark & \checkmark & $\kappa d$ & $D\log n$ & $\log n$ \\
  \hline
 \end{tabular}
 
 \begin{tablenotes} 
 \item [\textasteriskcentered] `orig:'  the original value of the property in the graph (i.e. the value in the graph $\G'$ in our model)\\
\end{tablenotes} 
\end{threeparttable}

\begin{threeparttable}[b]
\begin{tabular}{|l|p{0.14\textwidth}|p{0.23\textwidth}|p{0.1\textwidth}|p{0.08\textwidth}|p{0.08\textwidth}|}
\hline
 & \multicolumn{3}{|c|}{Costs} &  \multicolumn{2}{|c|}{ }\\
 \hline
 & Repair time & \# Msgs  per deletion& Msg size & match lower bound \tnote{$\ddagger$} & locality (hops) \tnote{$\sharp$} \\
 \hline
 DASH & $O(\log n)$ \tnote{$\dagger$}& $O(\delta\log n + \log^2 n)$ \tnote{$\dagger$}& $O(\log n)$ & \checkmark  & 1  \\
 \hline
 Forgiving Tree\tnote{!} &$O(1)$& $O(\delta)$ &$O(\log n)$ & \checkmark & 2 \\
 \hline
  Forgiving Graph\tnote{!} & $O(\log \delta \log n)$ &$O(\delta \log n)$ & $O(\log^2 n)$& \checkmark & $\log n$  \\
  \hline
Xheal/Xheal+ & $O(\log n)$  & $O(\kappa\log n A(p)/p)$ \tnote{\$}& $O(\log^2 n)$& \checkmark & $\log n$  \\
  \hline
 \end{tabular}
 
 \begin{tablenotes} 
 \item[$\dagger$] \emph{with high probability},  and amortized over $O(n)$ deletions.
 \item[$\ddagger$] The lower bounds differ according to the properties being bounded.
 \item[$\sharp$] Number of hops from the deleted node to nodes involved in repair.
 \item[\$] Amortized over $p$ deletions. $A(p)$ is a function of $p$, and $k$ is a parameter.
 \item[!] Algorithms using Virtual graphs
\end{tablenotes} 
\end{threeparttable}

\caption{Comparison of our self-healing Algorithms. $d$ is the degree of an individual node,  $\Delta$ is the maximum degree of a node in the graph, and $\delta$ is the degree of the deleted node.}
\label{tab: AlgoCompare}
\end{table}

\begin{figure}[h!]
\centering
\includegraphics[scale=0.65]{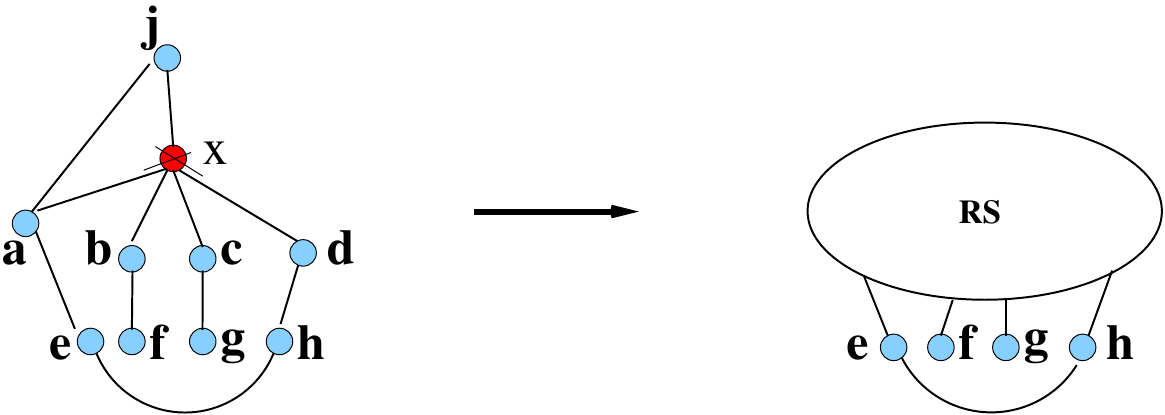}
\caption{Deleted node $x$ (in red, crossed) replaced by a Reconstruction Structure(RS), a structure formed by its neighbors ($a, b, c, d, j$). }
 \label{fig: RSheal}
\end{figure}

Table~\ref{tab: AlgoCompare} compares our self-healing algorithms in this line of work: DASH~\cite{SaiaTrehanIPDPS08}, ForgivingTree~\cite{HayesPODC08},  ForgivingGraph~\cite{FG-DCJournal2012,HayesPODC09}, Xheal/Xheal+~\cite{PanduranganPODC11, SarmaTrehanEdgeNetSciComm2012}.

Conceptually, Our algorithms use the same basic idea: when a node is deleted, replace it by a healing structure formed from its neighbors or nearby nodes, as shown in Figure~\ref{fig: RSheal}. We can call this structure  the $\RStr$. Notice that we have defined Reconstruction Structures like a template and the exact structure is determined  according to the desired properties of the algorithm. In most of our algorithms (DASH, ForgivingTree, ForgivingGraph ) the healing structure is a tree. In DASH, a balanced binary tree of neighbors with nodes arranged by previous degree increases is used. In the ForgivingTree,
 another kind of binary balanced tree is used.  In the Forgiving Graph
our reconstruction tree is a \textit{haft}(or half-full tree),
 In Xheal, the structure used is an expander.

 It turns out that in most of  these, trees are a natural choice for the graph properties we have tried to maintain. A balanced tree is a structure which has low distance between nodes (at most $2 \log_2 n$ for a balanced binary tree) while each node has a small degree  (at most 3 for a binary tree). At the same time, coming up with the suitable $\RT$s and maintaining them over the run of the algorithm is quite a significant challenge. For Xheal, however, trees are not the right structure since the main property we are trying to heal here is edge expansion and such spectral properties, and trees do not have good edge expansion.


\section{The idea of Virtual Graphs}
\label{sec: virtualgraphs}

\begin{figure}[h!]
\centering
\includegraphics[scale=0.5]{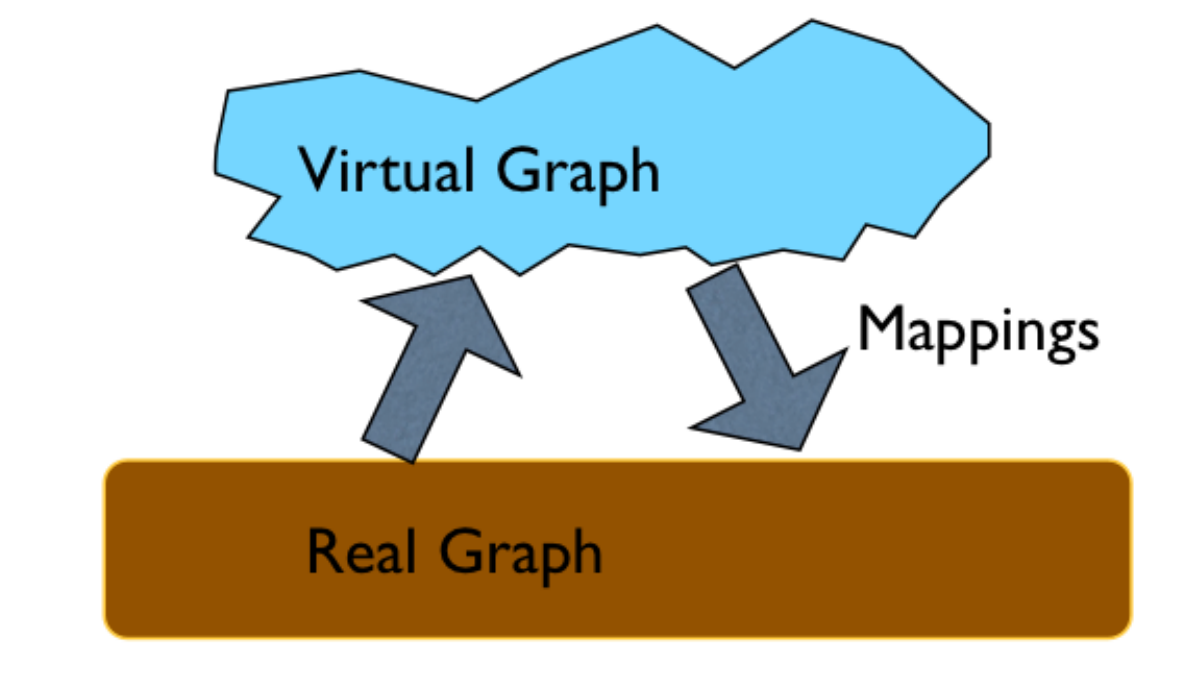}
\caption{Virtual Graph:  a mix of 'real' and 'virtual' nodes with appropriate mappings to the underlying real graph. }
 \label{fig: VRmap}
\end{figure}

An idea that we have sometimes found very useful is the idea of using \emph{virtual nodes}. A virtual node can be thought of as a marker or a placeholder in a reconstruction structure. A virtual node will be \emph{simulated} by a real node (we call the existing non virtual nodes as real nodes). Informally, simulating would simply mean that the simulating node takes responsibilities of the connections attributed to the virtual node (more formally discussed later)
In our algorithms, a virtual node is simulated by exactly one real node, but it may be possible to imagine algorithms where one virtual node may be simulated by multiple real nodes. Of course, one may have a single real node simulating multiple virtual nodes.  In our algorithms, the resulting graph that we maintain is a mixture of real and virtual nodes. We call this a \emph{virtual graph}. This is  opposed to the \emph{real graph}, which is the usual bijective mapping of the network to the graph with a processor mapping to a node and a connection mapping to an edge.  Section~\ref{sec: homomorphism} discusses a simple mapping used in the ForgivingTree and the ForgivingGraph that gives the real graph from the virtual graph and also shows some simple properties helpful for bounding certain properties.

 More formally,  consider the actual graph $G(V,E)$ corresponding to the network, and a virtual graph $G'(V',E')$ with nodes $V'$ and edges $E'$. Consider  a partition of the set $V'$ into two sets $V'_1$ and $V'_2$ (possibly empty), with  a surjective mapping $f: V'_2 \rightarrow V$ and a  mapping $g: V'_1 \rightarrow V'_2$, 
The edge sets $E'$ and $E$ are related by a mapping, a natural mapping being the homomorphism given in Section~\ref{sec: homomorphism}, possibly other mappings maybe imagined. Then, we have:
\begin{description}
\item{\bf real node:} A node $v' \in V'_2$.  By definition, we have a node $v \in V$, such that $f(v') = v$.
\item{\bf real edge:} An edge $(v',w') \in E'$ such that both $v'$ and $w'$ are real nodes.
\item{\bf virtual node:} Node $w' \in V'_1$. By definition, we have a node $v' = g(w')$, where $v' \in V'_2$. We say the real node $v'$ \emph{simulates} virtual node $w'$.
\item{\bf virtual edge:} An edge $(v',w') \in E'$ such that not both $v'$ and $w'$ are real nodes.
\end{description}


Virtual graphs are useful for a few conceptual reasons, some of which are:
\begin{itemize}
\item Virtual graphs may be easier to analyze and are good accountability structures (e.g. for bounding node degrees or distances). For example, if our Reconstruction Structure is a tree, and the virtual node is one of the internal nodes, we can claim that the node simulating it has increased its degree by at most 3 due to that virtual node.
\item A well designed virtual graph scheme may give a clearer insight into the structure and workings of the algorithm i.e. virtual graphs may be easier to visualize. Sometimes, the real graph and its connections may look messy and the underlying pattern, if any, may be obscured. However, this may be alleviated by the virtual graph. For example, the ForgivingTree data structure is a virtual graph that is a tree (and the algorithm is a tree maintenance algorithm) whereas the real graph corresponding to the ForgivingTree may not be a tree at all.

\item Virtual structures are easier to manipulate. After all, virtual nodes and edges are not really there, so algorithmically, it could be easy to drop or add them. Also, they act like placeholders for the real node simulating them and it is easy to imagine preserving the virtual structure while changing the ownership of the node, thus, manipulating the real structure. This is especially useful when reasoning about dynamic strucutres.
\end{itemize}

\begin{figure}[h!]
\centering
\includegraphics[scale=0.5]{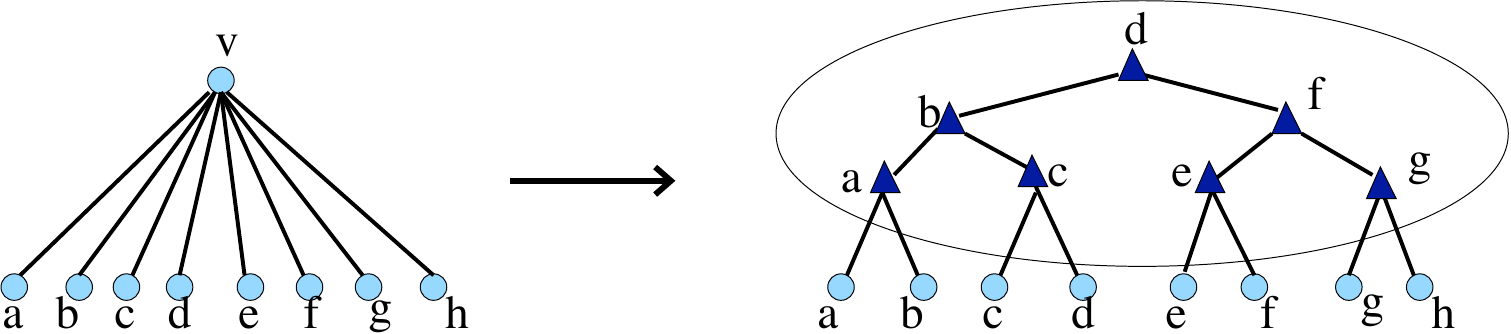}
\caption{Deleted node $v$ replaced by its Reconstruction Tree. The  triangle shaped nodes are 'virtual' helper nodes simulated by the 'real' nodes which are in the leaf layer.}
 \label{fig: RT}
\end{figure}

The Forgiving Tree and Forgiving Graph have reconstruction structures which use virtual nodes, and these structures are also trees. We call these reconstruction structures Reconstruction Trees and define them as follows:\\
\noindent{\bf Reconstruction Tree:}
\label{def: RTree} A tree like structure (Figure~\ref{fig: RT}) added by the healing algorithm on adversarial deletion of a single node and its edges. The reconstruction tree uses existing real nodes  from the network and may also may have virtual nodes simulated by the real nodes. 

\subsection{A framework for using virtual graphs}
 How to use virtual graphs for self-healing? As before, let $G$  be the real graph. Let $\mathcal{A}$ be the self-healing algorithm and $\mathcal{X}$ be the desired set of invariants on $G$. Let $I(G,\mathcal{A,X})$ be the  condition that $\mathcal{A}$ successfully maintains $\mathcal{X}$ on graph $G$ i.e. the self-healing algorithm is successful. A method for successfully using a  virtual graph for self-healing will be to come up with such a virtual graph and necessary mappings such that if the algorithm successfully maintains some invariants on the virtual graph, this implies success of the algorithm on the real graph. More formally, we need to develop an algorithm $\mathcal{A}$, virtual graph $G'$, set of invariants $\mathcal{X'}$ on $G'$ such that $I(G',\mathcal{A,X'})$ implies $I(G,\mathcal{A,X})$. 

When to use virtual graphs? It would be useful to have an idea of what problems and structures could lend themselves to solutions (hopefully elegant) using virtual graphs. This would be somewhat akin to using transformations in Algebra to solve a problem in a different basis. For example, the approach highlighted previously for self-healing could be extended to any algorithmic problem on the right structures. However, even for our own self-healing algorithms, we have found some problems which seem to be amenable to solution using virtual graphs (e.g. ForgivingTree, ForgivingGraph) and some otherwise (DASH, Xheal).

Virtual graphs beyond self-healing? There may be many problems and problem domains where it may be useful to  use a virtual graphs framework. A general framework could be on the lines above: create an appropriate virtual graph with real and virtual nodes with suitable mappings/reduction. Reduce the problem on the real graph to the virtual graph and solve it on the virtual graph so that the results hold for the real graph. For example, in certain problems involving mobile computing and mobile networks, it may be useful to have virtual nodes as placeholders (e.g. fixed positions where an agent is always present).

\subsection{De-simulation: Real Graph from a virtual Graph}
\label{sec: homomorphism}

\begin{figure}[h!]
\centering
\includegraphics[scale=0.5]{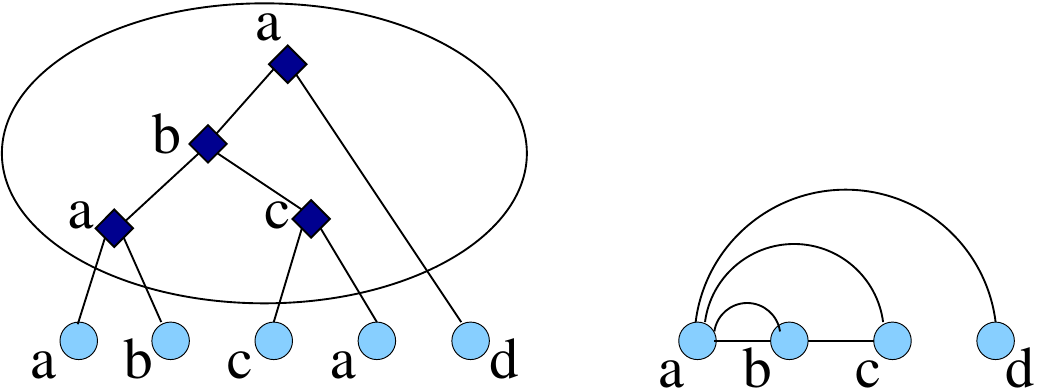}
\caption{The actual graph (on the right) is a homomorphic image of a virtual graph (left) where the helper nodes are mapped to the nodes simulating them. Note both the node degrees and distances between nodes in the real graph cannot be more than those in the virtual Graph.}
\label{fig: homomorphism}
\end{figure}

In the ForgivingTree/ForgivingGraph, a virtual graph maps to a real graph in a straightforward way: map all the virtual nodes to the real nodes simulating them.  Figure~\ref{fig: homomorphism} shows an example. 
More formally, the real graph is a homomorphic image of the virtual graph. Consider two graphs $G' = (V', E')$, and $G = (V, E)$.
In this context, a homomorphism may be defined as follows: 
A homomorphism is a function $f: V' \rightarrow V$ such that if undirected edge $\{v,w\}$  is in $E'$ (the edge set of $G'$) this implies that the edge $\{f(v), f(w) \}$ is  in $G$. Moreover, we say that $G$ is the homomorphic image of $G'$ under $f$ if the edges of $G$ are exactly the images of the edges of $G'$ under the homomorphism.
There can be multiple real and virtual nodes corresponding to a processor in the network  that perform all the functions required of those nodes. Each node can be identified by its processor and some additional information. For node $v$ in $G'$, let $Processor(v)$ be  the name of that processor. In the real graph $G$, there is only one node per processor and consider this node to be labelled with the name of that processor. Then,  our homomorphism $H:V(G') \rightarrow V(G)$ is simply $H(v) = Processor(v)$. 

As mentioned earier, we need to show that $I(G',\mathcal{A,X'})$ implies $I(G,\mathcal{A,X})$. Since the ForgivingGraph/ForgivingTree give bounds on distances and degree increase, the observations below suffice to show the required bounds in the papers (we refer the reader to the paper for  details, but intuitively, bounding the number of virtual nodes  simulated in the virtual graph by a real node bounds the real node's degree, if the virtual nodes are of constant degree).

\begin{observation}
\label{obs: homodegree}
For any graph homomorphism $F: G' \rightarrow G$, for all nodes $u, v$ in $V$,
  $dist_{G}(F(u), F(v)) \le dist_{G'}(u,v)$ where $dist_{G}(x,y)$
   is the distance between two nodes $x$ and $y$ in a graph $G$. 
\end{observation}

\begin{observation}
\label{obs: homostretch}
If the graph $G$ is the homomorphic image of  graph $G'$ under a graph homomorphism $F: G' \rightarrow G$,  then for all nodes $v'$ in $G$, $deg_{G}(v') \le \sum_{v \in F^{-1}(v')} deg_{G'}(v)$, where $deg_{G}(x)$ is the degree of the node $x$ in a graph $G$.
\end{observation}




\section{Directions and Conclusions}
\label{sec: conc}
This paper discussed self-healing in dynamic networks and introduced a responsive and scalable approach towards self-healing in reconfigurable networks.  A general model of self-healing and recent algorithms on self-healing using this and similar models were succinctly compared. A general idea in these algorithms is to replace a deleted node by a Reconstruction Structure. We also introduce a virtual graph framework and  a generic idea of using virtual structures (which may be useful for many problems besides self-healing) was introduced. We contend that this approach may be useful for a wide range of problems.


\bibliography{selfheal} 
\bibliographystyle{plain}

\end{document}